\documentclass{article}
\usepackage{hyperref}
\usepackage{natbib}
\setcitestyle{numbers}
\setcitestyle{square}
\usepackage{amsmath}
\usepackage{amssymb}
\usepackage{amsthm}
\usepackage{mathtools}
\usepackage{braket}
\usepackage{bm}
\usepackage{bbm}
\usepackage{graphicx}
\usepackage{xcolor}
\DeclarePairedDelimiter{\abs}{\lvert}{\rvert}
\newcommand{\numberset}{\mathbb}

\newcommand{\Z}{\numberset{Z}}
\newcommand{\C}{\numberset{C}}

\newcommand{\sigv}{\sigma_\text{v}}

\newcommand{\hsigd}{{\sigma}_\text{d}}

\newcommand{\ri}{i}
\newcommand{\bC}{\mathbf{C}}
\newcommand{\bCo}{\mathbf{\bar{C}}}

\newcommand{\bG}{\mathbf{G}}

\title{
\large \bf Reconstructive martensitic phase transitions: \\
intermittency, anti-trasformation, plasticity, irreversibility}
\date{\small \today}
\author{{\normalsize Edoardo Arbib$^1$, Noemi Barrera$^1$}, \\{\normalsize Paolo Biscari$^1$, Giovanni Zanzotto$^2$}\thanks{email: \href{mailto:edoardo.arbib@mail.polimi.it}{edoardo.arbib@mail.polimi.it},
\href{mailto:noemi.barrera@polimi.it}{noemi.barrera@polimi.it},
\href{mailto:paolo.biscari@polimi.it}{paolo.biscari@polimi.it}, \href{mailto:giovanni.zanzotto@unipd.it}{giovanni.zanzotto@unipd.it}}\\ \ \\ \small $^1$ Department of Physics, Politecnico di Milano, Italy\\
\small $^2$ DPG, Universit\`a di Padova, Italy
}

\begin{document}

\maketitle

\noindent \null \hfill {\it To Dick James, long-time friend,}\\
\noindent \null \hfill {\it whose great depth never impeded the great fun}

\bigskip
\smallskip

\begin{abstract}

We study the mechanics of temperature-driven reconstructive martensitic transformations in crystalline materials, within the framework of nonlinear elasticity theory. We focus on the prototypical case of the square-hexagonal  transition in 2D crystals, using a modular Ericksen-Landau-type strain energy whose infinite and discrete invariance group originates from the full symmetry of the underlying lattice.
In the simulation of quasi-static thermally-driven transitions we confirm the role of
the valley-floor network in establishing the strain-field transition-pathways on the symmetry-moulded strain energy landscape of the crystal. We also observe the phase change to progress through abrupt microstructure reorganization via  strain avalanching under the slow driving. We reveal at the same time the presence of assisting anti-transformation activity, which locally goes against the overall transition course. Both transformation and anti-transformation avalanches exhibit Gutenberg-Richter-like heavy-tailed size statistics. A parallel analysis shows agreement of these numerical results with their counterparts in empirical observations on temperature-induced martensitic transformations. The simulation furthermore shows that, in the present case of a reconstructive transformation, strain avalanching mostly involves lattice-invariant shears (LIS). As a consequence, microstructure evolution is accompanied by slip-induced defect nucleation and movement in the lattice. LIS activity also leads to the development of polycrystal grain-like lattice-homogeneity domains exhibiting high boundary segmentation in the body. All these effects ultimately lead to transformation irreversibility.

\bigskip
\bigskip

\end{abstract}

\section{Introduction}

We explore in this paper some main phenomena related to reconstructive martensitic transformations (RMTs) in crystalline materials, which is a widespread class of solid-to-solid phase transformations wherein the point-group symmetries of the parent and product phases obey no group-subgroup relation \citep{toledanoreconstructive, ContiZanzotto, BCZZnature}. RMTs include most famously, for simple Bravais lattices, the bcc-fcc structural transition taking place in many metals, alloys, and minerals, of great importance in science and technology \citep{booktransformationsmetals}.
We focus on the case of RMTs driven by temperature, using a non-linear elasto-plastic model in the 2-dimensional (2D) case, following Ericksen's early proposal of an infinite, discrete invariance group for the material's strain energy \citep{eri77, eri80}. This reflects the full symmetry of the underlying lattice, and collects all the deformations that map the lattice onto itself.
For Bravais lattices, this invariance is given by (a conjugate) of the group \(\text{GL}(n,\Z)\),
of unimodular (thus invertible) $n$ by $n$ matrices with integral entries ($n$ = 2, 3), see \cite{ContiZanzotto, BCZZnature, eri77, eri80, folkins,  parry, PZbook, bhattabook, pacoreview, jamesreview}. These produce, by group action, countably many  symmetry-related ground states of the material in strain space. For brevity we refer to this assumption as to the GL-invariance or modularity of the energy.

In the last decades, this far-reaching extension of the Landau theory was used in 3D, together with direct methods in variational calculus, obtaining much insight into symmetry-breaking martensitic transformations (MTs), with the associated phenomena of twinning and fine microstructure formation  \citep{PZbook, bhattabook, jamesreview, balljames1, balljames2, dolzmannbook, jamesperspective1, stefanmuller, 2013james_nat, jamesperspective2}.
In these MTs lattice strains are finite but small enough, i.e. confined within suitable 'Ericksen-Pitteri neighborhoods' (EPNs) in strain space, wherein global lattice symmetry reduces to local point-group symmetry \citep{ContiZanzotto, BCZZnature, eri80, PZbook, jamesreview, pitterireconciliation}. This reduces the relevant ground states (phase variants) to a finite number, and relates to the reversibility of the structural transformation \citep{jamesreview, jamesperspective1, 2013james_nat, jamesperspective2, acklandniti}, the crystal remaining largely defect-free.

However, a wide range of important phenomena in crystal mechanics originates from large strains reaching and going beyond EPN boundaries,
as in crystal plasticity, and in the reconstructive transformations presently studied.
For the purpose of studying the latter, a growing literature has explored, primarily in the 2D case, the consequences of GL-models allowing for large strains, unconfined to any EPNs \citep{ContiZanzotto, BCZZnature, pacoreview, pacoreview3, biurzaza, PRLgruppone, levslip}.

A specific line of research in this area has used explicit GL-invariant potentials developed in \citep{IJP, Jelasticity} for the 2D case, based on the proposal \citep{parry} of using the 'Klein modular invariant' $J$ \citep{modularforms2, modularforms1, mathematicaJ, wikiJ}. This plays the role of a global 2D-extension of the 'transcendental order parameter' in \cite{transcendental1, transcendental2}. See also \citep{folkins, vancouver} for other modular-functions approaches to 2D-crystal energetics.\footnote{A formulation of the present Ericksen-Landau GL-energy framework for crystals in 3D can be envisaged based on results in \citep{vancouver, BCCFCC3D, BCCFCC3D1, BCCFCC3D2}. Also, explicit fundamental domains in 3D, as are described in \cite{grenier, engel, terrasvol2, schwarzenberger, michel}, can be used to extend energy functions by GL-periodicity in 3D (see also \citep{parry3D}), as done explicitly in 2D with polynomials in \cite{ContiZanzotto}. A GL-informed phase-field model for crystal plasticity in 3D was studied in \cite{biurzaza}.}

In the present work we use a simplest class of GL-potentials proposed in \cite{Jelasticity} for the investigation of the prototypical RMT in two dimensions, i.e. the square-hexagonal ($s$-$h$) transformation in 2D Bravais lattices. This is of interest per se \cite{toledanoreconstructive, ContiZanzotto, squarehexagonal1, squarehexagonal2}, as well as being a highly instructive 2D proving ground for its 3D counterparts mentioned above. While in \cite{Jelasticity} a strain control was considered for the $s$-$h$ RMT, we use here the same constitutive function in a numerical quasi-static test driven by temperature. This produces a number of interesting results regarding the phenomena associated with such RMTs. Firstly, also in the present case, as with crystal plasticity \cite{IJP} and mechanically-driven RMTs \cite{Jelasticity}, the role is confirmed of the energy-surface valley-floors as transformation pathways for the evolving strain field. Phase- and defect-microstructure evolution thus depends not only on the  lattice-induced GL-arrangement, and kinematic compatibility, of the ground states in strain space, but also on the GL-topology of the valley floors connecting the minimizers on the energy surface, in particular on their bifurcations and possible loops \cite{IJP}.

Another main observation concerns the intermittent progress of the phase-micro\-structure via strain avalanching, which we find to follow Gutenberg-Richter-type heavy tailed statistics in the present 2D GL-model.
This concurs with previous indications of criticality effects in phase-transforming crystals, both numerical  \cite{pacoreview, crossover1, pacoreview2}, and experimental \citep{barcelona1, crossover2, barcelona3, PRBbarrera, barcelona2, drivingdirection}.

As a related novel effect, we report here the presence of systematic anti-trans\-for\-ma\-tion avalanching, locally running counter the transition course imposed by the external driving. This parallel strain activity possibly assists and promotes the leading transformation bursts in better achieving energy minimization and stress relaxation. Also the anti-transformation is found to exhibit heavy-tailed avalanche size statistics, expectedly more rapidly decaying than those pertaining to the on-course transformation avalanches.

An analysis, which we specifically performed for this study, of data desumed from empirical observations \cite{2013james_nat} of MTs induced by temperature, shows remarkable agreement of the above numerical results with their counterparts in real materials. This holds for the overall avalanche morphologies, as well as for their size statistics, which we have separately determined, in the numerical as well as in the empirical datasets, for both the transformation and the anti-transformation events.

Finally, our simulations also confirm that in the reconstructive case the evolving strain field spreads through neighboring EPNs in strain space.
As the spontaneous strains of the RMT reach the EPN boundaries, the global GL-invariance of the energy creates pathways to the ensuing activation of large lattice-invariant shears (LIS) in both the coexisting phases. The RMT structural change thereby forces ever increasing LIS activity during the transformation bursts experienced by the driven crystal. On the one hand, when LIS is highly localized, lattice-slip occurs, with the associated generation of dislocations, with LIS-layering especially inducing lattice-defect accumulation. Plastification thus typically accompanies RMT microstructure evolution in the crystal, as was noted in previous literature \citep{ContiZanzotto, BCZZnature, pacoreview, pacoreview3}.

Extended LIS-domains also occur in the simulation, creating, together with other LIS-microstructures, pseudo-twinned zones of LIS-reconstructed lattice homogeneity. This result from the model goes along with the well-established observation of polycrystal-like texturing produced in materials undergoing RMTs \citep{BCZZnature, natureirreversibility, orientationproliferation, laguna1, irreversibilitycompression2D}.

All the above effects in our simulation, whose common origin lies in RMT-strains forcing EPN exploration in strain space, contribute to defect accumulation with order degradation in the crystal, with the ensuing irreversibility of the structural transformation, a distinguishing feature of
RMT mechanics.

\section{Kinematics and energetics of phase-trans\-form\-ing 2D crystals on the upper complex half plane}

We consider a hyperelastic crystal in 2D, whose strain-energy density $\sigma$, due to Galilean invariance, is expressed as a constitutive function $\sigma = \sigma(\bC)$, where $\bC=
\bC^T>0$ is the symmetric, positive-definite Cauchy-Green strain tensor of the deformation field. This function must satisfy the material symmetry requirement:
\begin{equation}
\sigma(\bC)=\sigma(\bG^T\bC\bG),
\label{eq:energyinvariance}
\end{equation}
for any $\bG$ belonging to a suitable tensor group $\cal G$. As mentioned in the Introduction, for a crystalline material based on 2D Bravais lattices, we consider $\cal G$ to coincide with (a conjugate of) $\text{GL}(2,\Z)$. Besides GL-invariance we also assume
\begin{equation}
 \sigma(\bC)=\hsigd(\bCo) + \sigv(\det\bC),
\label{eq:sigelastic}
\end{equation}
where $\hsigd$ represents a distortive term of the energy that depends on the unimodular tensor $\bCo=(\det\bC)^{-1/2}\bC$, while $\sigv$ is a convex penalty function accounting for the volumetric effects of the deformation. Specifically,  we set $\sigv(\det\bC) = \lambda(\det\bC-\log\det\bC)$, with large enough modulus $\lambda>0$ to keep volumetric effects small.

To derive explicit forms for the (non-convex) GL-invariant distortive component $\hsigd(\bCo)$ in (\refeq{eq:sigelastic}), it is convenient \citep{parry} to use the smooth bijection $z =\hat z(\bCo)$ that maps the 2-dimensional space of $2 \times 2$ unimodular (positive-definite, symmetric) strain tensors $\bCo$ onto the upper complex half-plane $\mathbb{H}$:
\begin{equation}
\hat z(\bCo)={\bar{C}}_{11}^{-1}({\bar{C}}_{12} + i) \in \mathbb{H},
\label{bijectionH}
\end{equation}
where
\begin{equation}
\mathbb{H} = \{ x+\ri y\in\C, y>0 \},
\qquad (ds)^2=\big[(dx)^2+(dy)^2\big]/y^2 \;,
\label{hyperbolicmetric}
\end{equation}
the latter being the standard hyperbolic metric on $\mathbb{H}$ \citep{poincarehalfplane2, poincarehalfplane1, poincarehalfplane3}, with $C_{ij}$ the components of the strain tensor $\bC$ in the reference basis.

The group $\text{GL}(2,\Z)$ acts isometrically on $\mathbb{H}$ through the (linear fractional) Moebius transformations with integral entries
\citep{parry, PRLgruppone, IJP}. The Dedekind tessellation in Fig.~1c \citep{dedekind1, dedekind5} shows geometrically this action on $\mathbb{H}$ by means of the GL-equivalent congruent copies of the fundamental domain
$\mathcal{D}=\{ z\in \mathbb{H} : \abs{z}\ge 1,\ 0\le\text{Re}(z)\le\tfrac{1}{2}\}$. The latter contains the unimodular strains corresponding to all possible ways a 2D Bravais lattice can be deformed, up to GL-symmetry. Points on the interior of $\mathcal{D}$ produce generic oblique lattices with no symmetry, while points on the boundary $\partial \mathcal{D}$ are associated with strains that generate lattices with non-trivial rectangular and rhombic symmetries. Finally, the corner
points $i$ and $\rho = e^{i\pi /3}$ respectively corresponding to a square and to a hexagonal lattice.
We remark that in the Dedekind tessellation, the maximal EPN \citep{ContiZanzotto} of the square configuration $i$ [of the hexagonal configuration $\rho$] is obtained by considering the union of the four [six] GL-copies of $\mathcal{D}$ around $i$ [around $\rho$]. Correspondingly, the fundamental LIS system for $i$ [$\rho$], leading to ground states at the centers of the neighboring EPNs, is given by the shear of one lattice spacing along along the densest crystallographic rows, which on $\mathbb{H}$ is the path $i \to (i\pm 1)$
[$\rho \to (\rho\pm1$)],
together with all its GL-related copies. These shears correspond to the primary-slip directions in the plasticity of square [hexagonal] lattices.

\begin{figure}
\centering
\includegraphics[height=9.5cm, width=9.5cm]{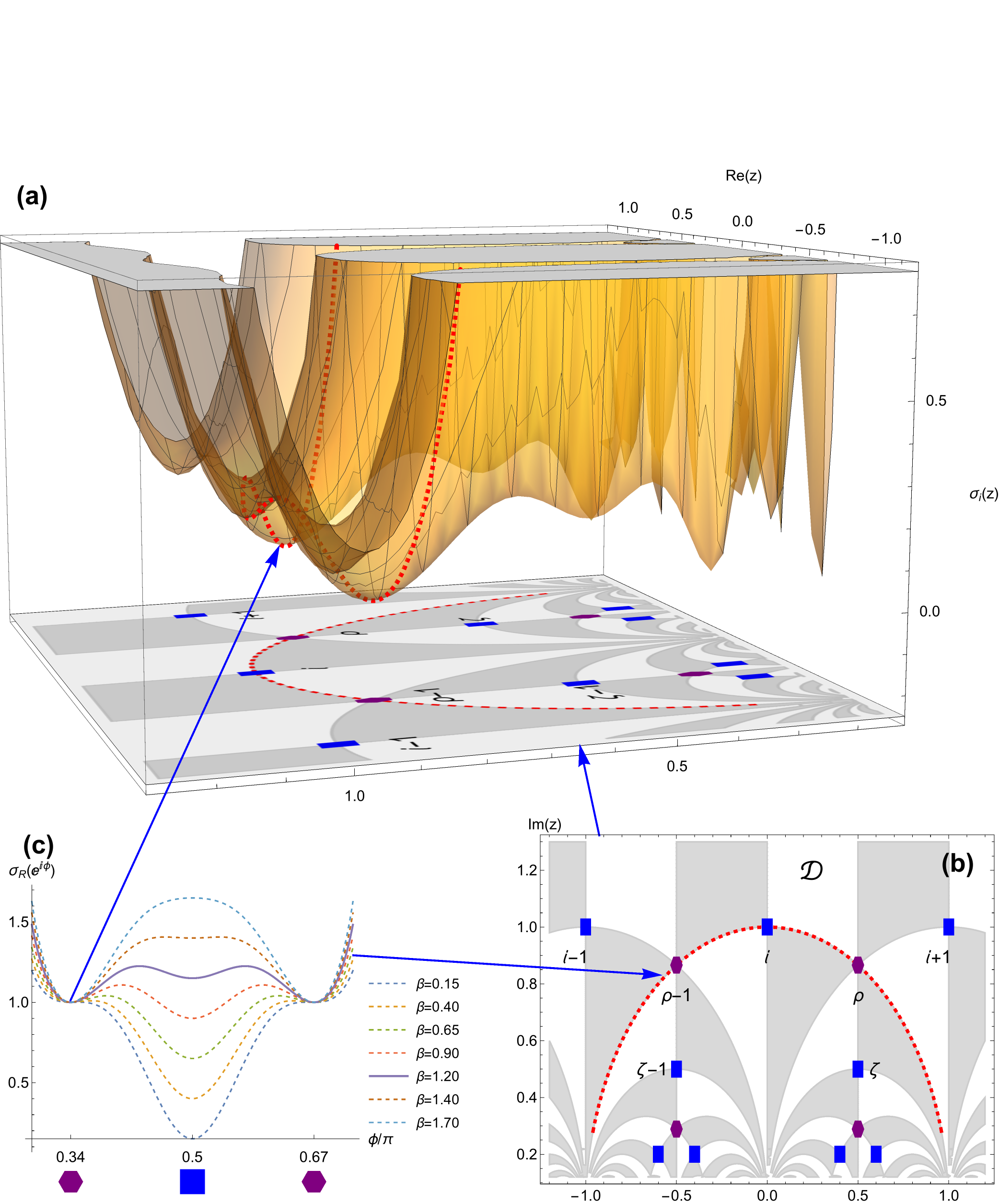}
\caption{\footnotesize (Color online) Ericksen-Landau strain energy landscape for the square-hexagonal ($s$-$h$) phase transformation. (a) Plot of the GL-invariant $s$-$h$ potential $\sigma_\text{d}$ in Eq. \eqref{eq:rec} for $\beta=1.2$ and $\mu=1$, displayed over a portion of the Dedekind tessellation on the hyperbolic plane $\mathbb{H}$.
(b) Dedekind tessellation of the hyperbolic Poincar\'e half-plane $\mathbb{H}$, illustrating the domain of the energy plot in panel (a). Gray and white regions denote GL-copies of the fundamental domain $\mathcal{D}$. Blue squares indicate the locations of nine GL-equivalent square points \big($i$, $i + 1$, $\zeta= \frac{1}{2}(i+1)$, $\zeta+1$, ...\big), while purple hexagons mark four GL-equivalent hexagonal points \big($\rho = e^{i \pi/3}$, $\rho -1$, ...\big). The GL-equivalent points appear closer together near the real axis, but remain equidistant in the hyperbolic metric (\ref{hyperbolicmetric})$_2$ on $\mathbb{H}$. The dashed red arc of the unit circle about the origin of $\mathbb{H}$ is a hyperbolic geodesic along which the $\beta$-dependent energy plots in panel (c) are obtained.
(c) Cross-sections for fixed $\beta$ of the $s$-$h$ energy surface $\sigma_\text{d}$ in \eqref{eq:rec}, taken along the unit arc (a hyperbolic geodesic), with phase $\phi$, shown as a dashed-red line in panels (a) and (b). The dashed energy profiles in (b) represent the cross sections of $\sigma_\text{d}$ for varying $\beta$, while the continuous line is the $\phi$-dependent cross section of the $\sigma_\text{d}$-surface depicted in panel (a), with $\beta = 1.2$.}
\label{fig:recons}
\label{fig:Dedekind}
\end{figure}

By means of the complex parametrization (\ref{bijectionH}) of strain space, useful families of GL-invariant Ericksen-Landau potentials can be derived by considering functions of the Klein invariant $J$ \citep{parry}, which is a well known $\text{SL}(2,\Z)$-periodic holomorphic function on $\mathbb{H}$ \citep{modularforms2, modularforms1, mathematicaJ,wikiJ}. Here $\text{SL}(2,\Z)$ denotes the positive-determinant subgroup of $\text{GL}(2,\Z)$. Thus, by defining $\hsigd(\bCo) =\hsigd\big(J(\hat{z}(\bCo)) \big)$, with suitable $\hsigd(J)$, we obtain smooth functions $\hsigd(\bCo)$ exhibiting the required full GL-periodicity. Furthermore, as $J$ is one-to-one from $\mathcal{D}$ to the closure of $\mathbb{H}$, the desired lattice configurations can be selected as ground states for these $J$-based  potentials \citep{PRLgruppone, IJP, Jelasticity, tesipatriarca}.

As mentioned in the Introduction, we consider here the most relevant 2D reconstructive transformation, involving the maximally-symmetric square ($s$) and hexagonal ($h$) lattice ground states. A simplest class of GL-invariant $J$-based potentials for the $s$-$h$ phase change was proposed in \cite{Jelasticity}:
\begin{equation}
    \sigma_\text{d}\big(\bCo\big) = \mu |J(z)-1|+\beta\mu |J(z)|^{2/3}  , \;\; \text{with}\ \; \mu>0 \;\text{ and } \; \beta>-\tfrac32,
    \label{eq:rec}
\end{equation}
where $z=\hat{z}(\bCo)$ as in (\ref{bijectionH}), and the elastic modulus $\mu$ is a scale factor, which we set to 1. With the bound (\ref{bijectionH})$_3$ on $\beta$, Eq. \eqref{eq:rec} defines a one-parameter family of GL-potentials $\sigma_\text{d}$ whose minimizers on $\mathbb{H}$ are given only by the GL-orbits of $i$ ($s$-orbit) and $\rho ={\rm e}^{i\pi/3}$ ($h$-orbit), see Fig.~\ref{fig:recons}a. The relative height of the corresponding minima is dictated by $\beta$ in \eqref{eq:rec} as shown in Fig.~\ref{fig:recons}c.
We refer to \cite{ContiZanzotto, IJP, Jelasticity} for a description of the GL-invariant $\beta$-bifurcation patterns for the $s$-$h$ critical points of $\sigma_\text{d}$. The parameter $\beta$ controls the driving in the quasi-static simulation of thermally induced $s$-$h$ transformations, as detailed below.

\section{Thermally-driven $s$-$h$ reconstructive \\ transformation}

We investigate numerically the behavior of an $h \to s$ phase-transforming crystal under quasi-static thermal loading ($10^4$ body cells for a total of $3\times 10^4$ degrees of freedom). We consider a typical metastable $s$-$h$ microstructure, as in Fig.~4d (obtained by quenching an initially homogeneous metastable square crystal), with initial value $\beta=1.4$, for which $h$ is the ground state. The control $\beta$ is then step-wise reduced through 100 equal decrements to the final value 0.5, for which the ground state is $s$ (Fig.~1c).
For each $\beta$, a strain field is determined by locally minimizing
the body's total strain-energy functional, computed from the density $\sigma_\text{d}$ in \eqref{eq:rec} (see details in \cite{IJP, Jelasticity}). Minimization is performed under minimal boundary conditions preventing inessential rigid-body roto-translations, specifically a hinge and a horizontal carriage respectively at the two ends of the lower side of the body, both friction-free.

\begin{figure}
\centering
\includegraphics[width=10cm]{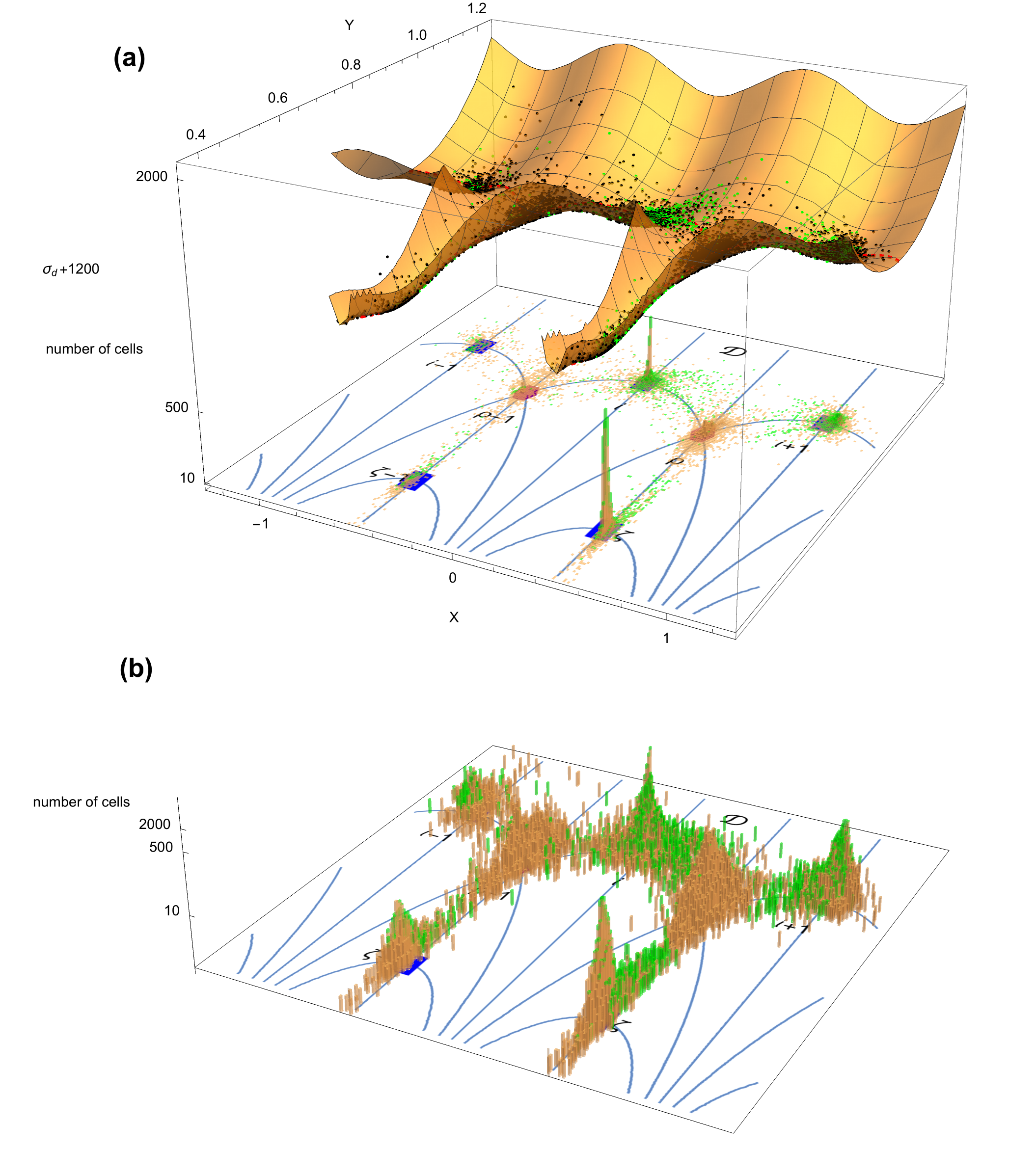}
\caption{\footnotesize{(Color online)
Snapshot illustrating the bursty evolution of the strain field over a portion of $\mathbb{H}$ during
a $\beta$-driven $h \rightarrow s$ transformation test ($\beta$ decreasing from 1.4 to 0.5). See the Supplementary Video \cite{video3} for the entire simulation. Panel (a) overlays the strain-field cloud
directly on the energy landscape defined on the Dedekind tessellation of Fig.~1(b). For each $\beta$, the unimodular strain tensor of each body cell
is represented by means of the bijection (\ref{bijectionH}) as a dot on the energy surface, given by the GL-energy density $\sigma_\text{d}$ in Eq.~(\ref{eq:rec}).
The snapshot shows the case for $\beta = 0.74$, with the $s$-wells (e.g. $i$, $i+1$, $\zeta$, \dots) deeper than the $h$-wells (e.g. $\rho$, $\rho-1$). During avalanches, the strain-cloud predominantly follows paths along the $\beta$-evolving energy valley-floors (indicated by dashed red lines on the surface), see \cite{IJP, Jelasticity}. Avalanching cells (i.e., cells jumping energy basin during the last imposed $\beta$-decrement) are marked as green dots on the surface, the rest are shown in black. A
large avalanche (in green) is highlighted in this snapshot.
Beneath the energy surface, a corresponding 2D histogram depicts the clustering of strain values evolving with $\beta$ on the Dedekind tessellation of $\mathbb{H}$.
The fraction of cells involved in the last strain avalanche is shown in green at the top of each bar. (b) Vertical log-scale representation of the same 2D histogram as in (a), highlighting the fraction of cell-strains elastically stabilized on the non-convex regions of the energy surface.
}}
\label{nubecanyon}

\end{figure}

\subsection{Strain avalanching: transformation intermittency and attending anti-transformation}

Our modelling captures many experimental aspects of martensitic transformations. One main effect, highlighted in this Section, is the intermittent progress of the phase microstructure, which is a known feature of these phase transitions (see the literature recalled in the Introduction).

\emph{Strain avalanching.}~~The avalanching transformation behavior in the simulation is clearly illustrated in Fig.~\ref{nubecanyon} and in the Supplementary Video \citep{video3}. We see there that, under the slow thermal driving, the basin-hopping of body cells on the GL-energy landscape proceeds in bursts, highlighted in green in the Figure and in the Video. As in previous studies of crystal plasticity \citep{PRLgruppone}, \citep{IJP}, and mechanically driven RMTs \citep{Jelasticity}, also in the present case we observe the basic role of the GL-network of energy valley floors as providing the low-energy pathways for the bursty evolution of the strain field during the phase change. They play a role
analogous, in our elasto-plastic setting, to the reaction pathways widely studied in other disciplines {\citep{monkeybiomolecules, saddleinflections2, saddleinflections5, proteinfolding, electronsmonkeysaddle}}.
GL-energy-topography analysis should prove useful also for informing other continuum-type approaches to crystal mechanics \cite{biurzaza, corridoifrancesi, denoualJMPSreconstructive, gao1, kan}, as valley-floor GL-topologies can be more complex \cite{PRLgruppone, IJP, tesipatriarca} than a simple network based on the GL-ground-state locations.

The 2D histograms on $\mathbb{H}$ in Fig.~2 and in the Video highlight that, during the minimization-driven transformation process,
the great majority of cell strains expectedly cluster near the energy-well bottoms in strain space  $\mathbb{H}$. The log-scale lowest 2D-histogram of Fig.~2(b) and in the Video \cite{video3}, however, emphasizes the fraction of cells' strains that are elastically stabilized on the non-convex regions of the landscape between wells. The presence of these marginally stable elements characterizes \cite{phaseorganization, socvaria1} the configurations of complex dissipative systems with many degrees of freedom and long range interactions, exhibiting cascade dynamics when slowly driven through a sequence of metastable states on a high-dimensional rough energy landscape.
Rheological example of these are the elasto-plastic crystals here considered, whose bumpy total-energy topography originates from the underlying global lattice symmetry of the strain energy \cite{pacoreview, pacoreview2, elliott1}. It is these marginally stable cells highlighted in Fig.~2(b) that are relatively more likely to trigger or participate to strain avalanching as $\beta$ varies, contributing in an essential way to the characteristic inhomogeneity of the MT processes, presently discussed.

Correspondingly, Fig.~3(a) (see also Fig.~4(g) below) shows the computed hexagonal phase fraction as a function of $\beta$, whose discontinuities originate from the combined effect of distinct strain avalanches producing local stress relaxation via spatially-separated microstructural rearrangements in the body. Examples of these strain avalanch\-es in the numerical simulations are illustrated in Figs.~3(b)-3(c).  At each $\beta$-step, they are identified as the connected components of the set of neighboring body cells whose strains move between two energy basins in a locally coordinated way. Multiple avalanches are thus mostly present at each $\beta$-decrement, whose \emph{size} $S$ is given by the total number of cells in each connected subset. Their aggregated effect determines the corresponding phase-fraction jumps in Fig.~3(a) and Fig.~4(g).

\begin{figure}

\centering
\includegraphics[width=6.1cm, height=5.5 cm]{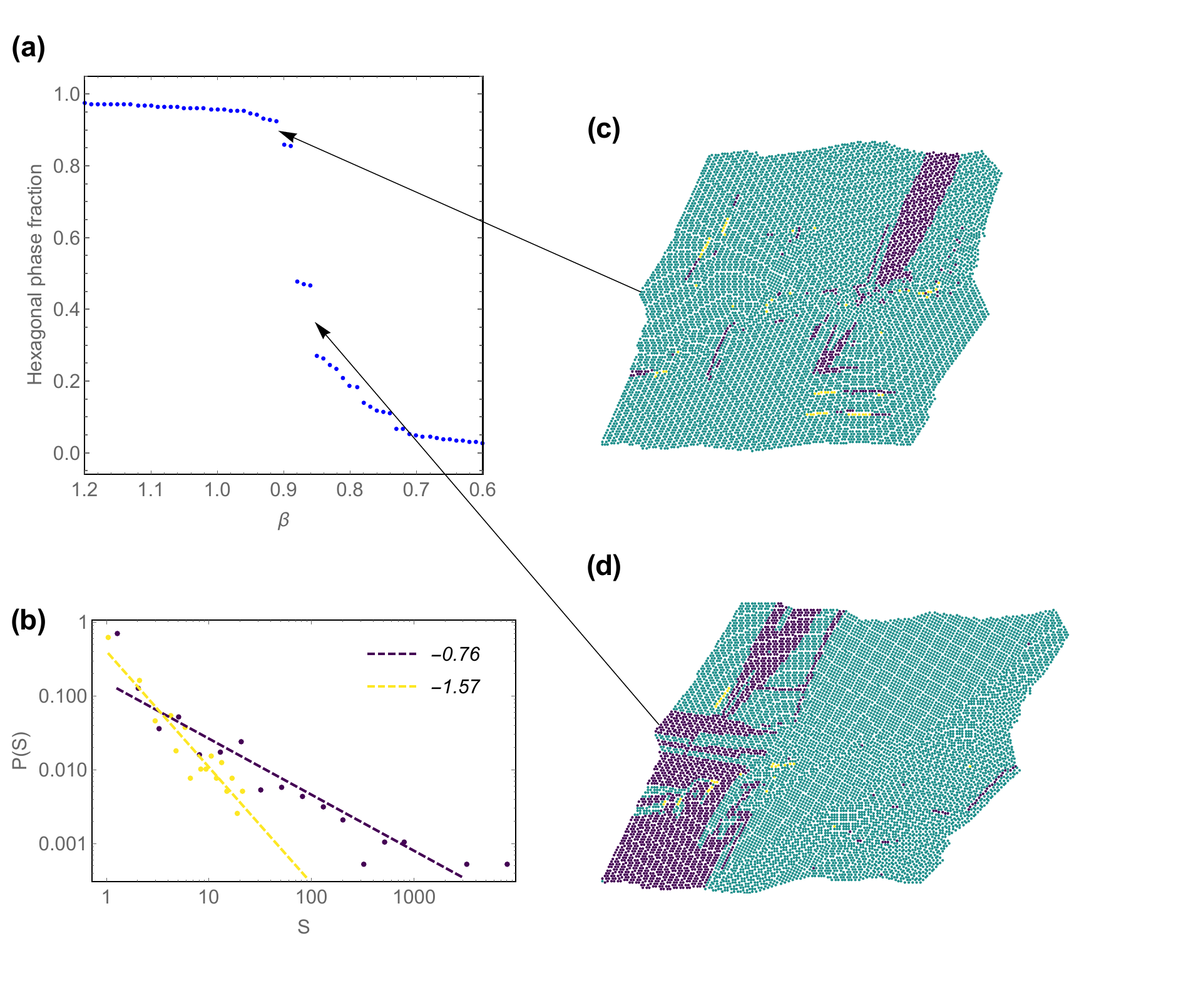}
\includegraphics[width=5.3cm, height=5.6 cm]{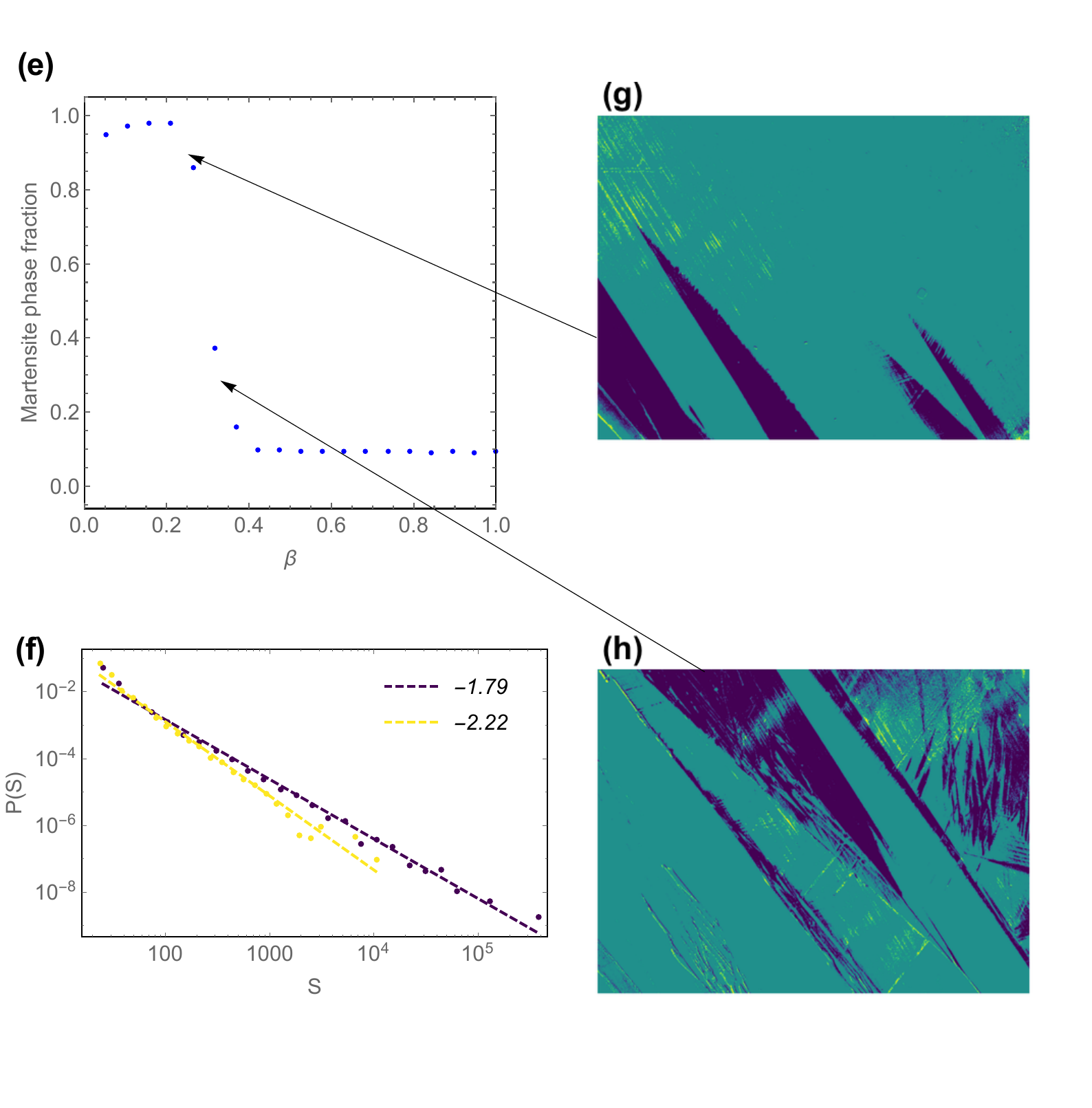}
\caption{\footnotesize
(Color online) Transformation intermittency due to strain avalanching under thermal driving.
Numerical simulation: (a)-(d); empirical observation: (e)-(h).
Panel (a) shows the computed discontinuous $\beta$-dependent hexagonal phase fraction in the $h \rightarrow s$ transforming body for decreasing $\beta$ (Fig.~2).
Snapshots (b) and (c) show examples of the transformation avalanches computed in the crystal, i.e.\ bursts of microstructural change in the simulation. The arrows indicate
the corresponding phase-fraction jumps. The color coding in (b), (c), distinguishes the
$h \rightarrow s$
transformation avalanches, in blue, vs. the smaller attending local anti-transformation
$s \rightarrow h$
events, in yellow. Panel (d) gives the associated log-log plots of the heavy-tailed probability density distributions $P(S)$ for the avalanche size $S$ (see text), separately for the transformation (blue) and anti-transformation (yellow) events in the simulation (the dashed lines, with the indicated slopes computed from least-squares fitting, are drawn to guide the eye).
Panel (e) shows the empirical phase fraction plot obtained from the analysis of the optically-recorded microstructures in a temperature-driven phase-transforming crystal from \cite{2013james_nat}. Snapshots in (f)-(g) show examples of transformation avalanches in the body during the thermally induced austenite-to-martensite phase change, with the same color coding as in (b)-(c).
Panel (h) gives, as in (d), the associated heavy-tailed avalanche-size distributions, separately for the observed transformation and anti-transformation events.}
\label{cdot}
\end{figure}

\emph{Local anti-transformation. } As a novel aspect in the present intermittency analysis, we distinguish here between the $h\to s$ transition avalanches, which align with the direction of the $\beta$-driving (blue domains in Figs.~3(b)-(c), and a minority of co-occurring $s\to h$ smaller events (yellow domains in the same panels) wherein cells move from an energy basin with lower value to one with higher value of the strain-energy density, indicating the presence of localized anti-transformation strain activity.
The latter appears to assist the primary transformation bursts during the evolution of the overall strain field, possibly enhancing stress relaxation in the course of total-energy minimization at each $\beta$-decrement. We count in this way a total of about 2,300 $\simeq$ 1,900 (blue) + 400 (yellow) avalanches during the simulation.

\emph{Comparison with empirical data. } The computational results in Figs.~3(a)-(b)-(c) can be compared with the MT in a real crystalline substance. We present in Figs.~3(e)-(f)-(g), the transformation avalanches derived from the analysis of optical empirical observations of a thermally driven phase-transforming metal alloy \cite{2013james_nat}. From these data, only a determination of the transformation activity on the body surface is possible, not the underlying strain intermittency within the driven crystal.\footnote{As the video data from \cite{2013james_nat} contain no strain information, to compare them in Fig.~3 with the simulation results we have directly counted, for each $\beta$-decrement, the local phase change produced by any cell-strain changing $h \to s$, or $s \to h$, energy basin. From the simulation data it is however possible to compute, beside the size, also the \emph{magnitude} of each strain avalanche, as the integral over its cell's strain-variations. This is in analogy to what was done in the experimental studies of strain intermittency through full-field surface strain measurement in stress-driven weak MTs in shape-memory alloys \cite{PRBbarrera, concurrent}, where both the size and magnitude of the observed strain avalanches were determined.
\label{nostrainavalanche}} We thus computed the evolving phase microstructures in \cite{2013james_nat} via the spatial color distribution on the surface of the video-recorded sample. The transformation avalanches were determined there by first identifying the domain $A$ in the 3-dimensional standard CIELAB
color space \cite{colorimetry}, derived from the parameters of the recorded austenite color-shades. For each video frame, the set of pixels with color belonging to $A$ was then computed, together with the set of pixels whose color moves into or out of $A$ compared to the previous frame. This produced the evolving phase fraction in Fig.~3(e), the associated transformation (blue) [anti-transformation (yellow)] avalanches being defined as the connected subsets of pixels whose colors transition out of [into] $A$ in a given frame. A threshold of 20 pixels was applied as a noise floor for avalanche size, defined, similarly to the numerical case, as the total number of pixels associated with each event. This gave a total of about 7,200 $\simeq$ 3,600(blue) + 3,600(yellow) avalanches
\footnote{We see from this count that the examined alloy from \cite{2013james_nat} exhibits relatively more abundant anti-transformation activity than in the simulation, as can also be seen also from the slopes of the statistic in Figs.~3(d)-3(h). This is possibly due to the engineered nature of the materials in \cite{2013james_nat}, which display enhanced scaling effects likely owing to their designed ease in forming and changing stress-free phase microstructures, with avalanche sizes exploring \cite{edinburghposter} a particularly extended range of scales, as shown in Fig.~11 of \cite{jamesreview}.} in the analyzed temperature run. Examples of these events are presented as blue and yellow domains in Figs.~3f-3g.

\begin{figure}
\centering
\includegraphics[width=13.5cm]{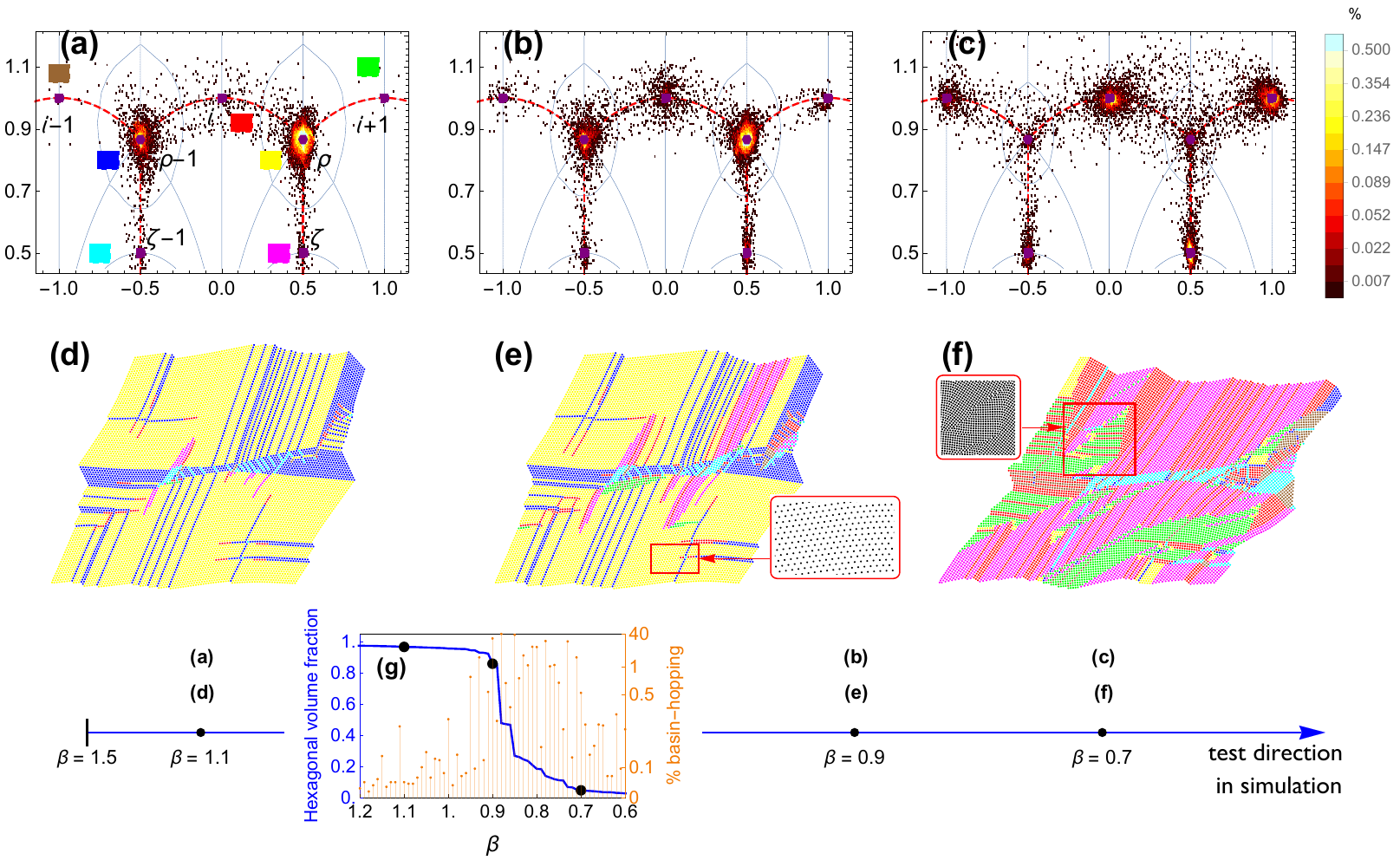}
\caption{\footnotesize (Color online) Bursty evolution of mixed-phase $s$-$h$ microstructure and associated strain-field cloud
on $\mathbb{H}$. The images refer to the thermally driven phase-transforming crystal, with gradually decreasing $\beta$ from $\beta = 1.4$ as described in Sect.~3. Panels (a),~(b),~(c) show three snapshots of the evolution of the $\beta$-dependent strain clustering during the test,
as a 2D heat-map histogram on the Dedekind tessellation of $\mathbb{H}$ in Fig.~1(b). See also the 2D strain histogram on the `floor' in Fig.~2; the Supplementary Video \citep{video3} shows the strain-cloud's evolution along the entire simulation. We see in these panels that, as in \cite{IJP, Jelasticity}, under the slow thermal driving the strain-cloud path on $\mathbb{H}$ follows the valley floors in the energy landscape (red dashed curves), crossing EPN-boundaries and visiting ever larger portions of strain space as the RMT progresses.
Panels (d),~(e),~(f) show the associated body deformation, characterized by the bursty evolution of the LIS-driven phase microstructure, with domains occupied by different variants of both the $h$ and $s$ lattice phases (color coding for the cell strains indicates the energy basin, as in panel (a)). For reference, panel (g) marks the position of the snapshots along the hexagonal phase fraction plot for decreasing $\beta$ (blue line), the same as in Fig.~3(a). The discontinuity of this plot tracks a global aspect
of the deformation's intermittency, indicated also by the orange spikes in the same panel, which show the bursty fraction of cell strains jumping energy basin at each $\beta$. Each burst is generated by a number of spatially separated strain avalanches in the body, as exemplified in Figs.~3(b)-(c). The corresponding evolution of the $s$-$h$ phase mixtures is accompanied by the creation and movement of lattice defects, as the dislocation in the detail inset to panel (e); see also Fig.~5(a). The inset to panel (f) shows one of the grain-like homogeneous-lattice domains produced by LIS activity, such as LIS-layering, as discussed in Sect.~3.2. }
\label{fig:snapshots}
\end{figure}

The agreement existing between the microstructure activity in our simulation and in the empirical observation can be seen in Figs.~3(b)-(c)-(f)-(g), in particular in the developing morphologies, largely governed by kinematic compatibility. We remark in both datasets the presence of the attending anti-transformation, demonstrated by the yellow domains in all the above figures.
To our knowledge this is the first time such phenomenon is highlighted in phase-transforming crystals, either as in the present computational prediction, or in empirical data.\footnote{Anti-transformation effects are also observed in the stress-driven uniaxial-extension experiments on martensite strain intermittency in \cite{PRBbarrera}, though this assisting activity is significantly less intense under the mechanical driving than with the present thermal case in Figs.~3(e)-(f)-(g) \cite{edinburghposter}.
}

The agreement between simulations and observations further extends to their statistical properties, as both exhibit  Gutenberg-Richter-type heavy-tailed avalanche-size distributions. This can be seen in Figs.~3(d)-3(h), providing additional
validation to the model.
The latter Figures suggest possible scaling behavior for the activity bursts, an observation aligning with previous modeling and experimental work reporting criticality effects in martensitic transformations, see the literature recalled in the Introduction.
this stage.

\subsection{Reconstructive-transformation irreversibility}
\label{irreversibility}

In the above Section we analyzed features pertaining to general martensitic transformations, which in particular also apply to RMTs. Hereafter, by contrast, we highlight effects which are specifically related to the reconstructive character of the MTs presently considered. They originate from the basic fact that, confirming the results in earlier investigations
\citep{ContiZanzotto, BCZZnature, pacoreview, pacoreview3}, the cell-strain values in our RMT-simulation systematically traverse the boundaries of multiple adjacent EPNs on the Dedekind atlas in $\mathbb{H}$, see Fig.~2, the Supplementary Video \cite{video3}, and Figs.~4(a)-(b)-(c).  As a consequence, in RMTs on the one hand systematic creation of dislocations occurs in the transforming lattice, and, on the other hand, the formation of polycrystal-type patterning. Both these RMT-accompanying phenomena have been recognized in past research on the irreversibility character of these structural transformations \citep{BCZZnature, natureirreversibility, laguna1, irreversibilitycompression2D}; see also the recent work \cite{levslip}.

\emph{Dislocations and local lattice disorder.}~~ One main consequence of EPN-visitation as a typical RMT-signature is that the corresponding phase microstructures (Figs.~4(d)-(e)-(f)) will largely be the result of lattice-invariant shears (LIS) for both the $s$ and $h$ coexisting phases. This is because at any site where the large spontaneous strain of the RMT or its symmetry-related variants occur, the local deformation, by definition,  gets near or across the local EPN boundary, and then on, along valley-floors, to the next minimizer in a neighboring well in strain space. In this way large energy-minimizing, lattice-reconstructing, shearing deformations, which belong to the GL-symmetry group of the lattice, become systematically activated, the barriers to plastification being here only as high as those to the phase change itself \cite{ContiZanzotto, BCZZnature}. The inset to Fig.~4(e) shows a detail of dislocation\footnote{The elastic field of lattice dislocations generated in the present GL-framework closely matches the predictions of the linear theory, see Fig.~1 in \cite{biurzaza}, Fig.~4 in \cite{PRLgruppone}.}
creation, as a highly localized LIS causes the slip of a lattice-line segment, dislocating the crystal at the end cell(s) bordering an unsheared segment on the same line \cite{ContiZanzotto, BCZZnature, pacoreview, pacoreview2}.

Fig.~5(a) highlights ($\beta = 0.9$) the dislocations so created by LIS activity throughout the phase-transforming crystal, evidenced as peaks in the spatial strain-energy distribution. In the Figure, energy thresholding heuristically separates the core zones from the background of elastically deformed cells, and we show in red on the 'ceiling' plane of Fig.~5(a) a snapshot of the resulting defect structures. We see plastification occurs with highest density at the borders of LIS layering, and in general at hetero-phase boundaries, as in Figs.~4(d)-(e)-(f). In the simulation this predominantly happens roughly along a main band across the body, where the most complex and fragmented LIS-generated $s$-$h$ microstructures first nucleate and then further develop through the nearby high stress regions.

The emergent defect structures themselves develop (Fig.~5(a)) in our framework through dislocation avalanch\-ing as a consequence of the underlying abrupt phase-microstructure evolution. Bursty defect creation and movement in the lattice, typical also of crystal plasticity \cite{zapperiDDDreview}, are thus obtained in this model as a by-product of cascading slip processes during energetics-driven strain bursts,
with no need for auxiliary hypotheses.  Analogous dislocational activity derived from strain avalanching in GL-type models of plasticity was investigated in \cite{biurzaza, PRLgruppone, IJP, comptesrendus}.

Besides LIS, further disorder sources in the lattice are the presence of valley-floor bifurcations in the energy topography, and possibly also the anti-transformation activity mentioned in the previous Section. Indeed, completion of any strain cycle via valley-floor loops \cite{IJP}, or via direct strain back-tracking in $\mathbb{H}$, might not always reverse the rotational part of a cell's deformation gradient, necessarily \cite{PZbook} creating additional disorder-inducing kinematical incompatibilities in the lattice. Again, this is the case also in crystal plasticity \cite{IJP}.

\emph{Grain-like homogeneity domains. } When LIS occurs either in layered patterns or in wider lattice bands as in Figs.~4(d)-(e)-(f), the lattice is by definition homogeneously reconstructed in its original orientation across any local shear plane. In both phases therefore LIS generate locally undetectable pseudo-twins \cite{PZbook} in the deformed configuration.\footnote{This is unlike with twinning shears, which reconstruct the lattice at a different orientation through the local twin boundary \cite{PZbook, bhattabook}. We remark that besides LIS-related slip and pseudo-twinning mentioned above, also actual twinning does occur in crystalline solids during out-of-EPN deformation regimes \cite{PZbook, natureirreversibility, austenitetwins1, austenitetwins2, austenitetwins3, GZborn}. This requires a more complex analysis than the much
investigated EPN-twinned microstructures (as in \cite{PZbook, bhattabook, jamesreview, balljames2, dolzmannbook, stefanmuller, 2013james_nat, Balandraudzanzotto, xavierfivedomains}), because the involved extra-EPN ground states are not finitely many. A further difficulty comes from the possible activation of additional lattice degrees of freedom, requiring larger lattice cells, as happens with cell-doubling in the well-known bcc-hcp transformation. This might make inadequate a modeling approach based solely on Bravais lattices. For results in these directions see \cite{PZbook, GZborn, acton, crocker, ironshear1, ironshear2, proliferationtwinning, magnesiumdoubletwinning} and references therein.}
These in turn create or rearrange, at each transformation burst, the domains of lattice homogeneity, leading to patterns of LIS-traversed grain-like zones in the body, each characterized by a different local orientation. These homogeneous-lattice domains can be seen in Figs.~4(d)-(e)-(f), involving, in those figures, different color-coded bands at their interior, referring to different LIS produced by EPN-boundary crossing. An example of a grain-like homogeneous-lattice domains generated by LIS activity is shown in the inset to Fig. 4(f), traversed by the  color-coded underlying LIS layers involving different strain basins.
The progress of the transformation in this way leads to polycrystal-type texturing with orientational proliferation in the driven crystal, which are long-recognized features of RMT activity \citep{BCZZnature, natureirreversibility, laguna1, irreversibilitycompression2D, orientationproliferation}.

Figs.~4(d)-(e)-(f) also show that at the boundaries of the above homogeneous domains, dislocations accumulation is more likely, with elastic marginal stabilization of cells as mentioned in the preceding Section (Figs.~2(b)). This leads to nuclei of metastable-phase retention and to incomplete transformation during the RMT processes (see also the observation of analogous phenomena in  \cite{irreversibilitycompression2D}), further contributing to RMT irreversibility.

\begin{figure}
\centering
\includegraphics[width=9cm]{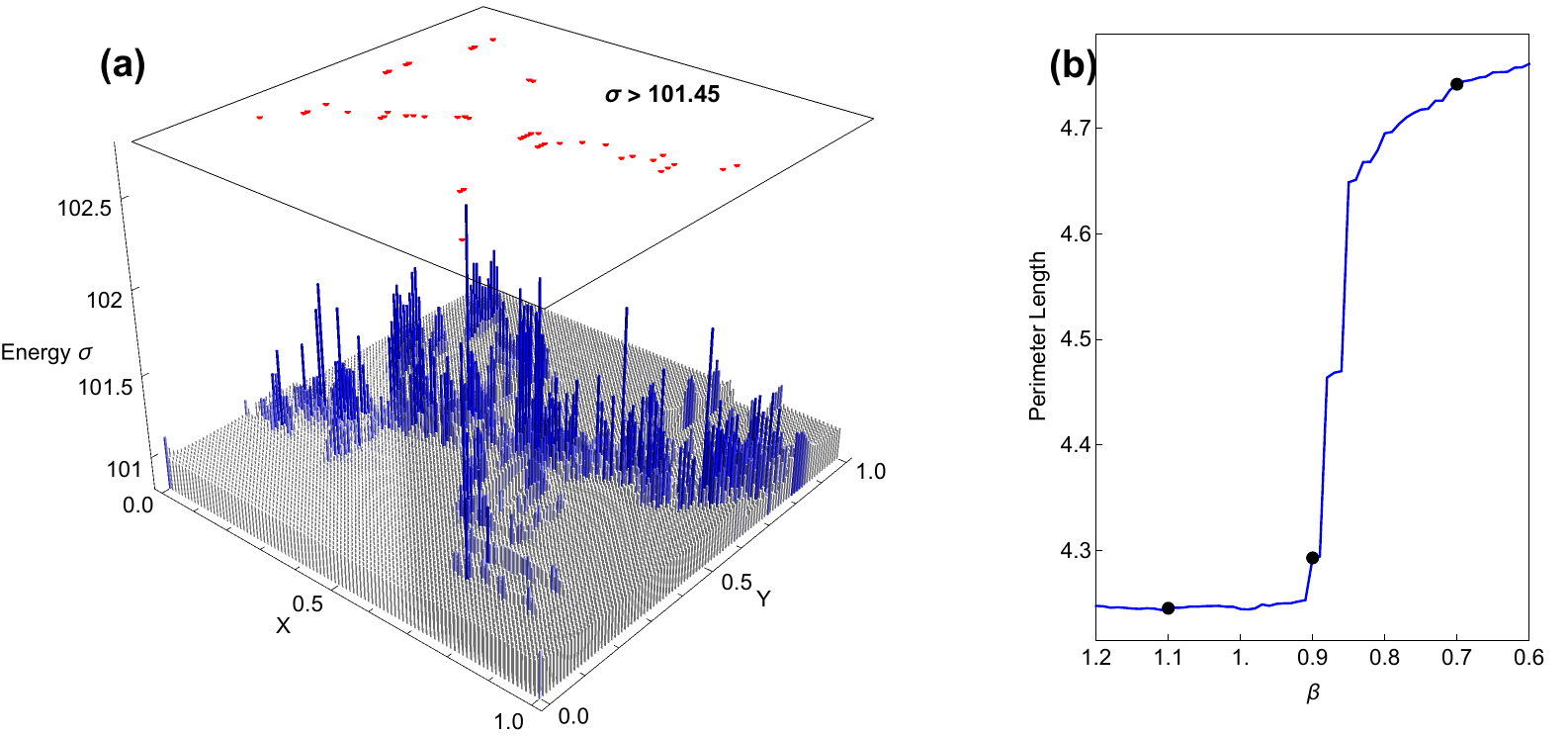}
\caption{\footnotesize (Color online) Irreversibility effects in thermally-driven
reconstructive transformations. (a) Snapshot for $\beta = 0.9$, showing a typical spatial distribution of the energy density on the body during the transformation (reference configuration). The peaks indicate the most dislocated lattice zones; on the the 'ceiling' plane are highlighted in red
the core regions by thresholding out the elastic background energy within the phase microstructure. These defects are created and evolve in the deforming body as a consequence of the strain avalanching induced by the thermal driving, see Figs.~4(d)-(e)-(f).
(b) Plot of the monotonic, intermittently increasing perimeter length of the deformed body (Figs.~4(d)-(e)-(f)) during the test. The black dots mark for reference the three $\beta$-values along the transformation process, as in Fig.~4(g).}
\end{figure}

\emph{Boundaries segmentation. } Yet another consequence of LIS activity observed in the simulation is the
growing boundary segmentation occurring in the homogeneity grains (Figs.~4(d)-(e)-(f)) within the transforming crystal. Fig.~5(b) shows this effect for the total body-perimeter in the deformed configuration. Its length is a monotonically increasing function of $\beta$, with discontinuities associated to the bursts of phase-microstructure change. It thus largely follows, but is not superposable to, the complementary to the discontinuous $h$-phase-fraction during the test, in Fig.~4(g).
Upon $\beta$ cycling the perimeter can be expected to grow monotonically as the boundary gets increasingly segmented by further LIS-accompanied microstructure evolution. This is because existing microstructures (as in Figs.~4(d)-(e)-(f)) have minimal chances of reversing their course as the body's strain-field cloud explores or cycles on different contiguous EPNs in $\mathbb{H}$ (Figs.~4(a)-(b)-(c)). Ever more pronounced boundary fragmentation occurs in the homogeneity-grains internal to the deforming body (Figs.~4(d)-(e)-(f)), due to the higher constraints imposed by the surrounding elasto-plastic matrix, compared to the entire free body referred to in Fig.~5(b). Fragmented grain boundaries within the lattice may turn into defect walls, with progressive loss of long-range lattice order and yet increased transformation irreversibility.

\section{Appendix: Caption to the Supplementary\\ Video~\cite{video3}}

This Supplementary Video \cite{video3} presents the quasi-static bursty evolution of the body's strain-field cloud on $\mathbb{H}$ in the simulation of a thermally-driven $s$-$h$ phase transformation. In the numerical test the parameter $\beta$ in the GL-energy density $\sigma_\text{d}$ in (\ref{eq:rec}) is decreased along the entire simulation (100 steps), from $\beta$ = 1.4, where the $h$ [$s$] configurations are ground lattice states [metastable states], to $\beta$ = 0.5, where the metastability of the $h$ and $s$ configurations is exchanged; see also the $\beta$-energy-profiles in Fig.~1(c).

(a) The visualization shows the evolving strain-cloud directly on the energy landscape along the $h \rightarrow s$ transformation. See a snapshot in Fig.~2. During the test, the strain values largely follow the energy valley floors indicated by dashed-red lines on the energy surface. See Fig.~4 for three snapshots of the deformed body configurations during the simulation, with the corresponding evolving $s$-$h$ phase microstructures. Cells undergoing avalanching, i.e., those changing energy well in the last $\beta$ decrement, are marked as green dots on the energy surface, with the remaining cells shown in black. The corresponding 2D histogram below the surface shows the clustering of strain values on $\mathbb{H}$ as it evolves with $\beta$. The fraction of cells participating in the last strain avalanche is represented in green at the top of each 2D-bin.

(b) The lowest 2D-histogram on $\mathbb{H}$ shows the same strain clustering as in (a) in vertical log scale, highlighting the fraction of cell strain values on the non-convex regions of the energy surface, elastically stabilized away from the well bottoms.

\bigskip

\noindent
{\it Aknowledgements. } We acknowledge the financial support of the Italian PRIN projects 2017KL4EF3 and  2020F3NCPX, and of INdAM-GNFM.

\end{document}